\documentclass[conference]{IEEEtran}
\usepackage{amsmath,amsopn,amssymb,bm}
\usepackage{amsfonts}
\usepackage{gensymb}
\usepackage{amsthm} 
\usepackage{graphicx,xspace,color}
\usepackage{epsfig}
\usepackage{longtable,multirow}
\usepackage{array}
\usepackage{stfloats}
\usepackage{cuted}
\usepackage{cite}
\usepackage[nolist]{acronym}
\usepackage{soul}
\usepackage{algorithm,algpseudocode}
\usepackage{algcompatible}
\usepackage{enumerate}
\usepackage{lettrine}
\usepackage{mathtools}
\usepackage{comment}
\usepackage{subcaption}

\usepackage[font=small]{caption}

\usepackage{booktabs}
\graphicspath{ {./figures/}}

\newtheorem{remark}{Remark}

\newtheorem{proposition}{Proposition}

\begin{acronym}
	\acro{THz}{terahertz}
	\acro{UPA}{uniform planar array}
	\acro{SW}{spherical wave}
	\acro{MIMO}{multiple-input multiple-output}
	\acro{IRS}{intelligent reflecting surface}
	\acro{IRSs}{intelligent reflecting surfaces}
	\acro{EM}{electromagnetic}
	\acro{MIMO}{multiple-input multiple-output}
	\acro{Tx}{transmitter}
	\acro{Rx}{receiver}
	\acro{RF}{radio-frequency}
	\acro{LoS}{line-of-sight}
	\acro{SNR}{signal-to-noise ratio}
	\acro{OFDM}{orthogonal frequency division multiplexing}
	\acro{EE}{energy efficiency}
	\acro{LoS}{line-of-sight}
\end{acronym}	

\IEEEoverridecommandlockouts 
\begin{document}
	
\title{Intelligent Reflecting Surface-Aided Wideband THz Communications: Modeling and Analysis}
\author{\IEEEauthorblockN{Konstantinos Dovelos\IEEEauthorrefmark{2}, Stylianos D. Assimonis\IEEEauthorrefmark{2}, Hien Quoc Ngo\IEEEauthorrefmark{2}, Boris Bellalta\IEEEauthorrefmark{1}, and Michail Matthaiou\IEEEauthorrefmark{2}}
	\IEEEauthorblockA{$^*$Department of Information and Communication Technologies, Universitat Pompeu Fabra (UPF), Barcelona, Spain}
	\IEEEauthorblockA{$^\dagger$Centre for Wireless Innovation (CWI), Queen's University Belfast, Belfast, U.K.}
	Email: \{k.dovelos, s.assimonis, hien.ngo, m.matthaiou\}@qub.ac.uk, boris.bellalta@upf.edu
}

\maketitle

\begin{abstract}
In this paper, we study the performance of wideband \ac{THz} communications assisted by an \ac{IRS}. Specifically, we first introduce a generalized channel model that is suitable for electrically large \ac{THz} \ac{IRS}s operating in the near-field. Unlike prior works, our channel model takes into account the spherical wavefront of the emitted electromagnetic waves and the spatial-wideband effect. We next show that conventional frequency-flat beamfocusing significantly reduces the power gain due to beam squint, and hence is highly suboptimal. More importantly, we analytically characterize this reduction when the spacing between adjacent reflecting elements is negligible, i.e., holographic reflecting surfaces. Numerical results corroborate our analysis and provide important insights into the design of future \ac{IRS}-aided \ac{THz}~systems. 
\end{abstract}

\begin{IEEEkeywords}
Beamfocusing, beam squint, intelligent reflecting surfaces, near-field, wideband THz communications. 
\end{IEEEkeywords}

\section{Introduction}
Unutilized spectrum resources are scarce in the sub-6 GHz band, which might limit the performance of future wireless communication systems. To this end, communication over the \acf{THz} band (0.1 to 10 \ac{THz}) is widely deemed a promising solution for beyond 5G networks due to the abundant spectrum available at those frequencies~\cite{6G_networks}. Despite the potential for terabit-per-second wireless links, \ac{THz} signals can suffer from severe propagation losses because of their short wavelength. Hence, transceivers with multiple antennas, i.e., \ac{MIMO}, are required to mitigate those propagation losses through sharp beamforming~\cite{prospects_multiantenna}. On the other hand, the power consumption of \ac{THz} radio-frequency circuits is much higher than their sub-6 GHz counterparts, which can undermine the deployment of massive antenna arrays in an energy efficient manner~\cite{thz_power_consumption}. To surmount this challenge, the novel paradigm of \ac{IRS}s can be exploited to reduce  the power consumption of a single-user \ac{MIMO} system, whilst fulfilling a transmission rate constraint~\cite{icc_paper}. Consequently, the performance analysis of \ac{IRS}-aided \ac{THz} communications is of great research importance. 

As shown in~\cite{icc_paper}, \ac{THz} \ac{IRSs} will likely operate in the radiating near-field where the wavefront of the emitted waves is spherical. In addition, the ultra-wide bandwidths, e.g., tens of gigahertz, of prospective \ac{THz} systems along with the large number of \ac{IRS} elements can yield a \textit{spatially wideband} channel~\cite{jsac_paper}. Under these circumstances, frequency-flat beamfocusing will decrease the power gain due to beam squint. To this end,~\cite{beam_squint_irs_ce, beam_squint_irs} have recently studied the
reflection design and channel estimation problem for \ac{IRS}-aided communications in the presence of beam squint, yet considering the far-field region. To the best of our knowledge, all prior studies on \ac{IRS}s (e.g.,~\cite{ref2,power_scaling_law_irs, ref3, ref4,ref5,ref6,irs_ofdm_design1,irs_ofdm_design2,irs_ofdm_design3,ref1}) neglect the spatial-wideband effect and/or the spherical wavefront of the radiated waves. This paper aims to fill this gap in the literature, and shed light on the channel modeling and performance of \ac{IRS}-assisted wideband \ac{THz} communications. We commence by introducing a spherical wave channel model for discrete \ac{IRSs}, which takes into account the spatial-wideband effect; note that our model includes the far-field regime as a special case. We next study the power gain under standard narrowband beamfocusing, and show that it is highly suboptimal. More importantly, we analytically evaluate the reduction in the power gain when the inter-element spacing is negligible, which corresponds to the ultimate limit of a discrete \ac{IRS} known as \textit{holographic reflecting surface}~\cite{ref1,holographic_mimo}. Our performance analysis reveals that beam squint mitigation through a frequency-selective surface design is essential for reaping the full potential of \ac{IRS}-assisted wideband \ac{THz} systems.
\begin{figure}[t]
	\centering
	\includegraphics[width=0.78\linewidth]{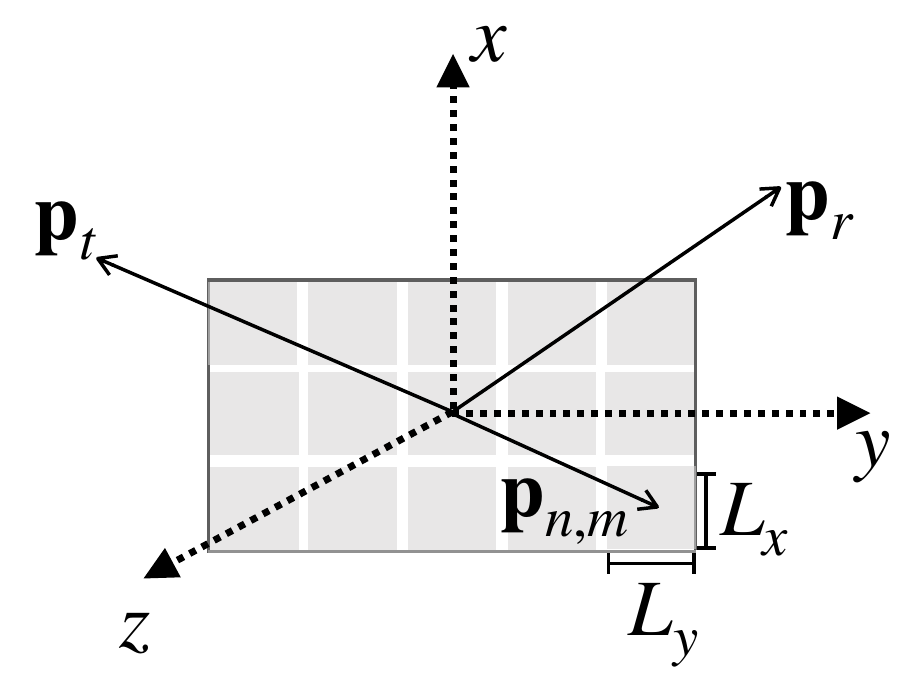}
	\caption{Illustration of the \ac{IRS} geometry under consideration.}
	\label{fig:irs_geometry}
\end{figure}

\textit{Notation}: Throughout the paper, $\mathbf{X}$ is a matrix; $\mathbf{x}$ is a vector; $x$ and $X$ are scalars; $(\cdot)^*$, $(\cdot)^T$, and $(\cdot)^H$ are the conjugate, transpose and conjugate transpose, respectively; $[\mathbf{X}]_{i,j}$ is the $(i,j)$th entry of $\mathbf{X}$; $\text{vec}(\mathbf{X})$ is the column vector formed by stacking the columns of $\mathbf{X}$; $\mathcal{F}\{\cdot\}$ is the continuous-time Fourier transform; $\mathbf{x}\sim\mathcal{CN}(\bm{\mu},\bm{\Sigma})$ is a complex Gaussian vector with mean $\bm{\mu}$ and covariance matrix $\bm{\Sigma}$; and $\text{erf}(x) = \frac{2}{\sqrt{\pi}} \int_{0}^xe^{-t^2}dt$ is the error function. 

\section{Channel Model with Spatial-Wideband Effects}
\subsection{System Setup}
We consider an \ac{IRS}-assisted \ac{THz} system, where the \ac{Tx} and \ac{Rx} have a single antenna each. The \ac{IRS} is placed in the $xy$-plane, and it consists of $N=N_x\times N_y$ passive reflecting elements of size $L_x\times L_y$ each, as depicted in Fig.~\ref{fig:irs_geometry}; the inter-element spacing is negligible and is ignored~\cite{ref6}. The origin of the coordinate system is placed at the center of the \ac{IRS}. The position of each  \ac{IRS} element is measured from its center. Then, the position vector of the $(n,m)$th \ac{IRS} element is $\mathbf{p}_{n,m}=((n-1/2)L_x,(m-1/2)L_y,0)$, for $n=-\frac{N_x}{2},\dots, \frac{N_x}{2}-1$, and $m=-\frac{N_y}{2},\dots, \frac{N_y}{2}-1$. Likewise, $\mathbf{p}_t = (r_t,\theta_t,\phi_t)$ and $\mathbf{p}_r = (r_r,\theta_r,\phi_r)$ are the position vectors of the \ac{Tx} and \ac{Rx}, respectively, where $r$ is the radial distance, $\theta$ is the polar angle, and $\phi$ is the azimuth angle. In Cartesian coordinates, the distance between the \ac{Tx} and the $(n,m)$th \ac{IRS} element is hence given by
\begin{align}\label{eq:tx_distance}
r_t&(n,m) \triangleq \|\mathbf{p}_t-\mathbf{p}_{n,m}\| \nonumber\\
& = r_t\left(1 + \frac{\left((n-\frac{1}{2})L_x\right)^2}{r_t^2} - \frac{2\cos\phi_t\sin\theta_t (n-\frac{1}{2})L_x}{r_t} \right . \nonumber\\
&   \left . + \frac{\left((m-\frac{1}{2})L_y\right)^2}{r_t^2} - \frac{2\sin\phi_t\sin\theta_t (m-\frac{1}{2})L_y}{r_t}\right)^{1/2}.
\end{align}
Similarly, the distance between the \ac{Rx} and the $(n,m)$th \ac{IRS} element is calculated as
\begin{align}\label{eq:rx_distance}
r_r&(n,m) \triangleq \|\mathbf{p}_r-\mathbf{p}_{n,m}\| \nonumber\\
& = r_r\left(1 + \frac{\left((n-\frac{1}{2})L_x\right)^2}{r_r^2} - \frac{2\cos\phi_r\sin\theta_r (n-\frac{1}{2})L_x}{r_r} \right . \nonumber\\
&   \left . + \frac{\left((m-\frac{1}{2})L_y\right)^2}{r_r^2} - \frac{2\sin\phi_r\sin\theta_r (m-\frac{1}{2})L_y}{r_r}\right)^{1/2}.
\end{align}
 
\subsection{Channel Model}
We focus on the \ac{Tx}-\ac{IRS}-\ac{Rx} link. The received baseband signal propagated through the \ac{IRS} is expressed as
\begin{equation}\label{eq:rx_signal}
r(t) = \sum_{n=-\frac{N_x}{2}}^{\frac{N_x}{2}-1}\sum_{m=-\frac{N_y}{2}}^{\frac{N_y}{2}-1} h_{n,m}e^{j\varphi_{n,m}} x(t-\tau_{n,m}) + \tilde{n}(t), 
\end{equation}
where $e^{j\varphi_{n,m}}, \varphi_{n,m}\in[-\pi, \pi]$, is the reflection coefficient of the $(n,m)$th \ac{IRS} element, $\tilde{n}(t)\sim\mathcal{CN}(0,\sigma^2)$ is the additive noise at the receive end, $x(t)$ is the transmitted baseband signal, and $\tau_{n,m}$ is the associated propagation delay given by
\begin{equation}\label{eq:total_prop_delay}
\tau_{n,m} = \frac{r_r(n,m) + r_t(n,m)}{c},
\end{equation}
where $c$ denotes the speed of light. Moreover,  
\begin{equation}
h_{n,m} = \sqrt{\text{PL}_{n,m}(f)} e^{-j 2\pi f_c \tau_{n,m}} 
\end{equation}
is the cascaded channel through the $(n,m)$th IRS element, $\text{PL}_{n,m}(f)$ is the corresponding frequency-dependent path loss, and $f_c$ is the carrier frequency. The path loss of the cascaded channel through the $(n,m)$th \ac{IRS} element is calculated as~\cite{icc_paper}
\begin{align}\label{eq:pl_expression}
&\text{PL}_{n,m}(f) = \nonumber \\ &G_tG_r\left(\frac{L_xL_y}{4\pi}\right)^2\frac{F(\theta_t,\phi_r,\theta_r)}{r_t^2(n,m)r_r^2(n,m)}e^{-\kappa_{\text{abs}}(f)(r_t(n,m) + r_r(n,m))},
\end{align}
where $F(\theta_t,\phi_r,\theta_r) \triangleq \cos^2\theta_t(\cos^2\theta_r\cos^2\phi_r + \sin^2\phi_r)$, $G_t$ and $G_r$ are the \ac{Tx} and \ac{Rx} antenna gains, respectively, while $\kappa_{\text{abs}}(f)$ denotes the molecular absorption coefficient at frequency $f$~\cite{thz_propagation_modeling}. Taking the Fourier transform of~\eqref{eq:rx_signal} gives
\begin{align}\label{eq:rx_signal_freq}
R(f) \approx & \sqrt{\text{PL}(f)}  \underbrace{\sum_{n=-\frac{N_x}{2}}^{\frac{N_x}{2}-1}\sum_{m=-\frac{N_y}{2}}^{\frac{N_y}{2}-1}\!\!\!e^{-j 2\pi (f_c +f)\tau_{n,m}}e^{j\varphi_{n,m}}}_{H_{\text{eff}}(f)} \! X(f) \nonumber\\
& + \tilde{N}(f),
\end{align} 
where the approximation follows from $\text{PL}_{n,m}(f)\approx \text{PL}(f)$ owing to the small physical size of \ac{THz} \ac{IRS}s~\cite{icc_paper}, $\text{PL}(f)$ denotes the path loss calculated using the radial distances $r_t$ and~$r_r$, $\mathcal{F}\{r(t)\} = R(f)$, $\mathcal{F}\{x(t)\} =X(f)$, $\mathcal{F}\{\tilde{n}(t)\} = \tilde{N}(f)$, and $H_{\text{eff}}(f)$ is the effective channel accounting for the phase shifts. Note that $H_{\text{eff}}(f)$ is frequency-dependent because of the spatial-wideband effect. Next, consider \ac{OFDM} modulation with $S$ subcarriers for a signal bandwidth $B$. The subcarrier spacing is given by $B/S$, and the baseband frequency of the $s$th subcarrier is specified as $f_s=\left(s-\frac{S-1}{2}\right)\frac{B}{S}$, for $s=0,\dots,S-1$. Thus, the received signal at the $s$th \ac{OFDM} subcarrier is given by 
\begin{equation}
R(f_s) =  \sqrt{\text{PL}(f_s)} H_{\text{eff}}(f_s) X(f_s) + \tilde{N}(f_s),
\end{equation}
where $X(f_s)\sim\mathcal{CN}(0,P_t/S)$ is the transmitted data symbol with average power $P_t/S$, and  $\tilde{N}(f_s)\sim\mathcal{CN}(0,\sigma^2B/S)$ is the additive noise at each subcarrier.
\begin{remark}[Fresnel Approximation]
In the radiating near-field, i.e., Fresnel zone, the \ac{Tx} distance can be approximated by $r_t(n,m)\approx r_t + \tilde{r}_t(n,m)$, where  
\begin{align}\label{eq:tx_approx}
\tilde{r}_t(n,m) &=\frac{\left((n-\frac{1}{2})L_x\right)^2(1-\cos^2\phi_t\sin^2\theta_t)}{2r_t}  \nonumber\\
&- \left(n-\frac{1}{2}\right)L_x\cos\phi_t \sin\theta_t \nonumber\\
&+  \frac{\left((m-\frac{1}{2})L_y\right)^2(1-\sin^2\phi_t\sin^2\theta_t)}{2r_t}  \nonumber\\
&- \left(m-\frac{1}{2}\right)L_y\sin\phi_t\sin\theta_t
\end{align}
follows from the second-order Taylor polynomial $(1+x)^\alpha \approx 1 + \alpha x + \frac{1}{2}\alpha(\alpha-1)x^2$ of \eqref{eq:tx_distance}. Similarly, it holds that $r_r(n,m)\approx r_r + \tilde{r}_r(n,m)$, where $\tilde{r}_r(n,m)$  is given by $\eqref{eq:tx_approx}$, but $\theta_t$, $\phi_t$, and $r_t$ are replaced by $\theta_r$, $\phi_r$, and $r_r$, respectively.
\end{remark} 

\section{Performance Analysis of \ac{IRS}-aided Wideband THz Communications}
\subsection{Power Gain}
From~\eqref{eq:rx_signal_freq}, the \ac{SNR} at the $s$th \ac{OFDM} subcarrier is written as  
\begin{equation}
\text{SNR}_s = \frac{N^2G_s P_t\text{PL}(f_s)}{B\sigma^2},
\end{equation}
where $G_s\in[0,1]$ is the normalized power gain defined as
\begin{equation}
G_s \triangleq \frac{\left| H_{\text{eff}}(f_s)\right|^2}{N^2}.
\end{equation}
With frequency-dependent beamfocusing, the phase induced by the $(n,m)$th \ac{IRS} element is $\varphi_{n,m}(f_s) = 2\pi (f_c + f_s)\tau_{n,m}$, which yields $G_s=1$ for each \ac{OFDM} subcarrier. Therefore, $\text{SNR}_s$ grows quadratically with the number $N$ of \ac{IRS} elements. Conversely, with conventional narrowband beamfocusing, we have $\varphi_{n,m} = 2\pi f_c\tau_{n,m}$ for all subcarriers, and 
\begin{align}\label{power_gain}
G_s = \frac{1}{N_x^2N^2_y}\left|  \sum_{n=-\frac{N_x}{2}}^{\frac{N_x}{2}-1}\sum_{m=-\frac{N_y}{2}}^{\frac{N_y}{2}-1} e^{-j 2\pi f_s\tau_{n,m}} \right|^2,
\end{align}
which results in $G_s < 1$ for $f_s >0$ due to beam squint.

We next generalize the normalized power gain to the case of a holographic \ac{IRS} modeled as a continuous aperture. To do so, we leverage the Fresnel approximations of $r_t(n,m)$ and $r_r(n,m)$ introduced in Remark 1. Then,~\eqref{power_gain} is recast as
\begin{equation}\label{eq:approx_gain}
G_s \approx \frac{1}{N_x^2N^2_y}\left|  \sum_{n=-\frac{N_x}{2}}^{\frac{N_x}{2}-1}\sum_{m=-\frac{N_y}{2}}^{\frac{N_y}{2}-1} \! e^{-j 2\pi f_s\frac{\tilde{r}_t(n,m) + \tilde{r}_r(n,m)}{c}} \right|^2.
\end{equation}
\begin{figure*}[t]
	\centering
	\begin{subfigure}{.5\textwidth}
		\centering
	\includegraphics[width=0.85\linewidth]{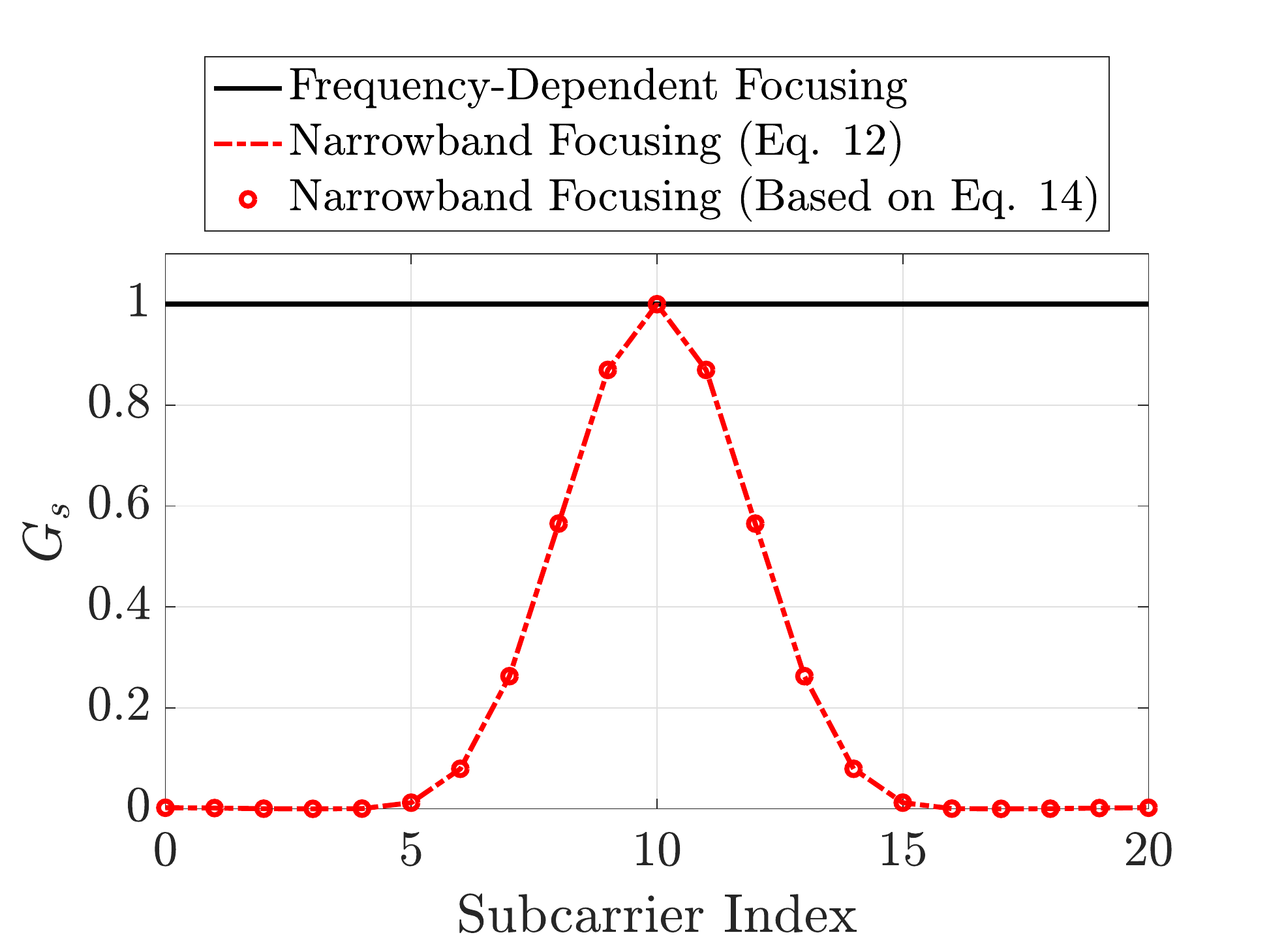}
		\caption{Normalized power gain}
		\label{Fig2a}
	\end{subfigure}%
	\begin{subfigure}{.5\textwidth}
		\centering
	\includegraphics[width=0.85\linewidth]{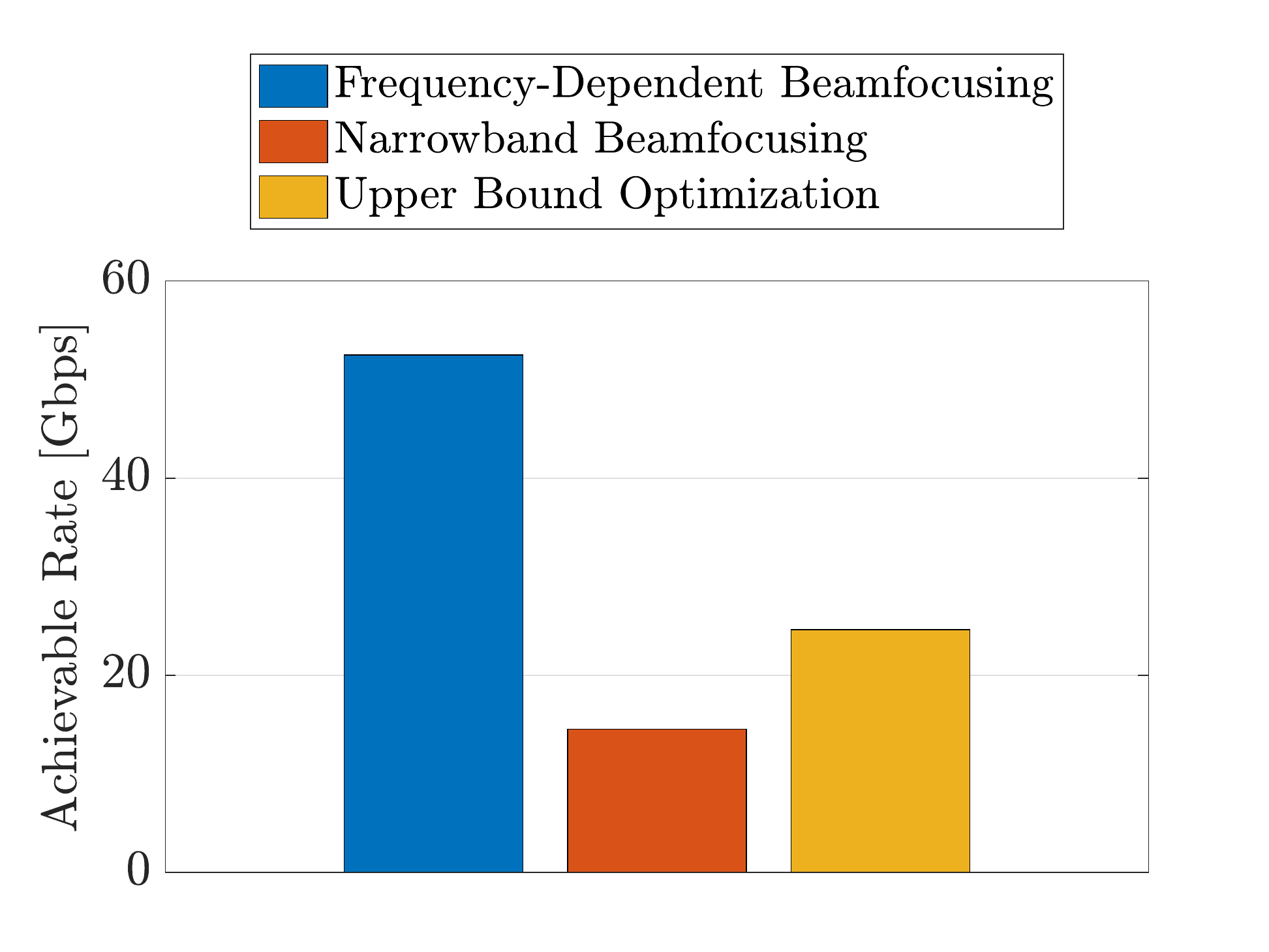}
		\caption{Achievable rate}
		\label{Fig2b}
	\end{subfigure}%
	\caption{Results for an $80\times 80$-element \ac{IRS}, $B=20$~GHz, $f_c = 300$ GHz, $(r_t,\theta_t,\phi_t) =  (1,\pi/3,\pi/5)$, and $(r_r,\theta_r,\phi_r) =  (5,\pi/4,\pi/3)$.}
	\label{Fig2}
\end{figure*}

In the following proposition, we provide an analytical expression for~\eqref{eq:approx_gain}.
\begin{proposition}[Holographic Reflecting Surface]
The normalized power gain at the $s$th \ac{OFDM} subcarrier is analytically evaluated as $G_s = |\xi_s|^2$, where $\xi_s$ is given by~\eqref{eq:cf_expression} at the top of the next page for $k_s = 2\pi f_s/c$, $\tilde{L}_x = N_xL_x$, $\tilde{L}_y = N_yL_y$, and 
\begin{align}
a_x  &=\frac{(1-\cos^2\phi_t\sin^2\theta_t)}{2r_t} + \frac{(1-\cos^2\phi_r\sin^2\theta_r)}{2r_r}, \tag{15}  \\
b_x & =\cos\phi_t \sin\theta_t + \cos\phi_r \sin\theta_r,  \tag{16} \\
a_y &= \frac{(1-\sin^2\phi_t\sin^2\theta_t)}{2r_t} + \frac{(1-\sin^2\phi_r\sin^2\theta_r)}{2r_r},   \tag{17}\\
b_y&=\sin\phi_t\sin\theta_t + \sin\phi_r\sin\theta_r.   \tag{18}
\end{align}
\begin{proof}
See Appendix.
\end{proof}
\end{proposition}
From Fig.~\ref{Fig2}(\subref{Fig2a}), we observe the detrimental effect of beam squint on the normalized power gain. Moreover, we verify the excellent match between~\eqref{power_gain} and the expression in Proposition~1, which implies that a spatially continuous \ac{IRS} can be accurately approximated by an ultra-dense discrete \ac{IRS}. 

\subsection{Achievable Rate via Beamfocusing Optimization}
To improve the system performance, we can resort to a more advanced \ac{IRS} design than narrowband beamfocusing. To this end, we introduce the auxiliary matrices $\mathbf{A}_s\in\mathbb{C}^{N_x\times N_y}$, with $[\mathbf{A}_s]_{n,m} = e^{-j2\pi f_s\tau_{n,m}}$, and $\mathbf{B}\in\mathbb{C}^{N_x\times N_y}$, with $[\mathbf{B}]_{n,m} = e^{j\varphi_{n,m}}$. Then, we have $|H_{\text{eff}}(f_s)|^2 = |\mathbf{h}^T_s\mathbf{b}|^2$, where $\mathbf{h}_s = \text{vec}(\mathbf{A}_s)\in\mathbb{C}^{N\times 1}$ and $\mathbf{b} = \text{vec}(\mathbf{B})\in\mathbb{C}^{N\times 1}$. We seek to find the reflection coefficient vector $\mathbf{b}$ that maximizes the achievable rate, i.e., 
\begin{equation}\label{eq:bf_opt_problem}
\begin{matrix}
\underset{\mathbf{b}}{\max} & R(\mathbf{b}) =   \sum_{s=0}^{S-1}\frac{B}{S}\log_2\left(1  + \frac{P_t\text{PL}(f_s)|\mathbf{h}^T_s\mathbf{b}|^2}{B\sigma^2}\right) \tag{19}\\[0.4cm]
\text{s.t.} & |[\mathbf{b}]_n|=1, \forall n=1,\dots, N. 
\end{matrix}
\end{equation}
The beamfocusing optimization problem \eqref{eq:bf_opt_problem} resembles the wideband design problem in \ac{IRS}-aided \ac{OFDM} systems, which is non-convex and difficult to solve~\cite{irs_ofdm_design3}. In the spirit of~\cite{irs_ofdm_design3}, we turn to maximize the upper bound\footnote{This optimization approach is well-established in the related literature; see~\cite{beam_squint_irs, irs_ofdm_design3,mmwave_squint}, and references therein.} of $R(\mathbf{b})$:
\begin{equation}
R(\mathbf{b}) \leq B\log_2 \left(1 +\frac{P_t\text{PL}(f_s)}{B\sigma^2} \frac{\sum_{s=0}^{S-1}|\mathbf{h}^T_s\mathbf{b}|^2}{S}\right).\tag{20}
\end{equation}
We therefore formulate the optimization problem
\begin{equation}
\max_{\mathbf{b}}  \quad \sum_{s=0}^{S-1}|\mathbf{h}^T_s\mathbf{b}|^2 = \|\mathbf{H}^T\mathbf{b}\|^2 = \mathbf{b}^H\mathbf{H}^*\mathbf{H}^T\mathbf{b},\tag{21}
\end{equation}
where $\mathbf{H} = [\mathbf{h}_0,\dots, \mathbf{h}_{S-1}]\in\mathbb{C}^{N\times S}$. The above quadratic form has the solution $\mathbf{b}^{\star} = \sqrt{N}\mathbf{u}$, where $\mathbf{u}$ is the unit-norm eigenvector corresponding to the maximum eigenvalue of the Hermitian matrix $\mathbf{H}^*\mathbf{H}^T$. Since the elements of $\mathbf{b}^{\star} $ do not satisfy the unit-modulus constraint, this solution is referred to as \textit{upper bound optimization}; recall that $\mathbf{b}^{\star}$ would be implemented by controlling the amplitude and phase of each reflection coefficient, which is not feasible in the passive \ac{IRS} architecture under consideration. As a result, the upper bound optimization serves as a benchmark to assess the impact of the spatial-wideband effect on the achievable rate. In Fig.~\ref{Fig2}(\subref{Fig2b}), the achievable rates of the frequency-dependent beamfocusing, narrowband beamfocusing, and upper bound optimization approach are 52.48~Gbps, 14.52~Gbps, and 24.61~Gbps, respectively. Narrowband beamfocusing performs very poor, and results in a $72.3\%$ rate loss. Moreover, the upper bound optimization approach performs better, yet yields a much smaller rate than frequency-dependent beamfocusing. This numerical experiment showcases the importance of having \ac{IRS} elements with a wideband response~\cite{fss}. 
\begin{figure*}[t]
	\begin{align}\label{eq:cf_expression}
	&\xi_s=\frac{\pi}{4jk_s\tilde{L}_x\tilde{L}_y\sqrt{a_xa_y}}\left[\text{erf}\left(\sqrt{jk_sa_x}\left(\frac{\tilde{L}_x}{2} - \frac{b_x}{2a_x}\right)\right)- \text{erf}\left(\sqrt{jk_sa_x}\left(-\frac{\tilde{L}_x}{2} - \frac{b_x}{2a_x}\right)\right) \right]\nonumber \\[0.1cm]
	&\quad\quad\quad\quad\quad \times  \left[\text{erf}\left(\sqrt{jk_sa_y}\left(\frac{\tilde{L}_y}{2} - \frac{b_y}{2a_y}\right)\right)- \text{erf}\left(\sqrt{jk_sa_y}\left(-\frac{\tilde{L}_y}{2} - \frac{b_y}{2a_y}\right)\right) \right].\tag{14}
	\end{align}
		\hrulefill
\end{figure*}  
\begin{figure*}[t]
	\centering
	\includegraphics[width=0.7\linewidth]{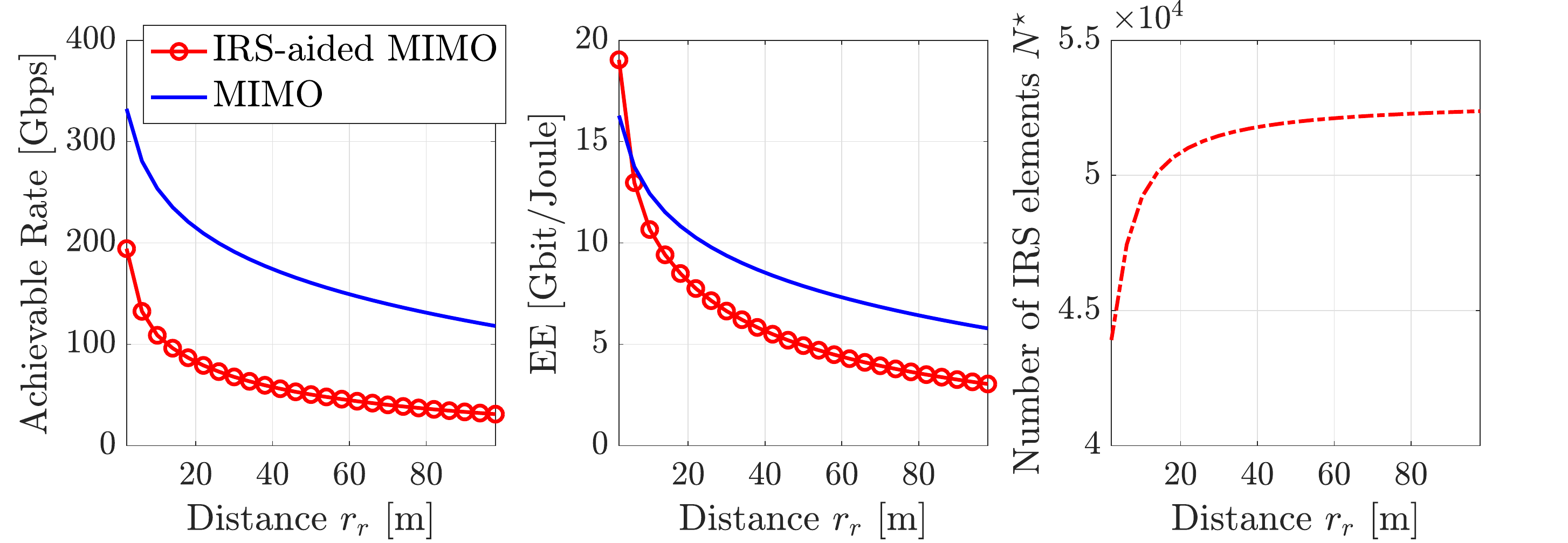}
	\caption{Results for $\alpha=2$, narrowband beamfocusing, and fixed IRS location at (0,0,0). In the MIMO system, $N_t = 100$ and $N_r = 100$. The other parameters are $G_t = G_r = 20$~dBi, $P_t = 10$ dBm, $\sigma^2 = -174$ dBm/Hz, $B=20$~GHz, $S=20$, $f_c = 300$ GHz, $L_x=L_y=\lambda/2$, $\mathbf{p}_t = (x_t,y_t,z_t) = (0.8,-0.8,0.2)$, and $\mathbf{p}_r = (x_r,y_r,z_r)= (0.8, r_r,0.2)$.}
	\label{fig:Fig_EE}
\end{figure*}
\subsection{Energy Efficiency}

\subsubsection{MIMO System}
Consider a \ac{MIMO} system, where the \ac{Tx} and \ac{Rx} have $N_t$ and $N_r$ antennans, respectively. For efficient hardware implementation, hybrid analog-digital array architectures are assumed at both ends. The frequency-dependent path loss of the direct channel, i.e., \ac{LoS}, is~\cite{jsac_paper}
\begin{equation}
\text{PL}_{\text{MIMO}}(f) = \frac{G_tG_r c^2}{(4\pi r_d (f_c+f))^2}e^{-\kappa_{\text{abs}}(f)r_d},\tag{22}
\end{equation}
where $r_d = \|\mathbf{p}_t-\mathbf{p}_r\|$. Next, we assume that $N_t$ and $N_r$ are adequately small so that the spatial-wideband effect is negligible; this can be attained by a uniform planar array~(UPA), such as an $10\times 10$-element UPA~\cite{jsac_paper}. In the far-field, the \ac{LoS} channel matrix is rank-one; this holds for all distances larger than the Fraunhofer distance $2D^2_m/\lambda$, where $D_m$ is the maximum dimension of the antenna array and $\lambda$ is the carrier wavelength~\cite[Ch.~7]{tse_book}. Then, frequency-flat beamforming and combining yield the received \ac{SNR}
\begin{equation}\tag{23}
\text{SNR}^{\text{MIMO}}_{s} = \frac{N_t N_r P_t \text{PL}_{\text{MIMO}}(f_s)}{B\sigma^2}.
\end{equation} 
The respective power consumption is calculated as\footnote{The power consumption of signal processing is neglected.}
\begin{equation}\label{power_cons_model}
P_{\text{MIMO}} = P_t + N_r(P_{\text{PS}} + P_{\text{PA}}) + N_t(P_{\text{PS}} + P_{\text{PA}} ),  \tag{24}
\end{equation}
where $P_{\text{PS}}$ and $P_{\text{PA}}$ are the power consumption values for a phase shifter and a power amplifier, which are 42~mW and 60~mW at $f_c = 300$~GHz, respectively~\cite{thz_power_consumption}.

\subsubsection{\ac{IRS}-Aided \ac{MIMO} System}
The \ac{Tx} and \ac{Rx} perform beamforming and combining to communicate a single stream through the \ac{IRS} of $N$ elements. Due to the directional transmissions, the \ac{Tx}-{Rx} link is very weak, and hence is neglected. In this case, the received \ac{SNR} at the $s$th \ac{OFDM} subcarrier is 
\begin{equation}
\text{SNR}_s = \frac{N_tN_r N^2G_s P_t\text{PL}(f_s)}{B\sigma^2}. \tag{25}
\end{equation}
Using varactor diodes, the power expenditure of an IRS element is nearly negligible~\cite{ref6}. For the sake of exposition, we assume that the power consumption of the \ac{IRS}-aided system is also given by~\eqref{power_cons_model}. Therefore, the \ac{EE} is given by $\sum_{s=0}^{S-1} \frac{B}{S}\log_2(1 + \text{SNR}_s)/P_{\text{MIMO}}$. Akin to~\cite{icc_paper}, we can now decrease the number of antennas as $N_t/\alpha$ and $N_r/\alpha$, whilst increasing the number of \ac{IRS} elements as $N^{\star} = \alpha\frac{\lambda}{L_xL_y}\frac{r_tr_r}{\sqrt{F(\theta_t,\phi_r,\theta_r)}r_d}e^{-\frac{1}{2}\kappa_{\text{abs}}(f)(r_d - r_r-r_t)}$, to attain an \ac{EE} gain~$\alpha$ compared to the pure \ac{MIMO} system. The achievable rate, \ac{EE}, and $N^{\star}$ are plotted versus $r_r$ in Fig.~\ref{fig:Fig_EE}. In contrast to the spatially narrowband case, \ac{IRS}-assisted \ac{MIMO} cannot outperform \ac{MIMO} because of the beam squint effect present in the former architecture. 
 
\section{Conclusions}
We have studied, for the first time, the spatial-wideband effect in \ac{IRS}-aided \ac{THz} communications. In particular, we introduced a spherical wave channel model that captures the peculiarities of wideband transmissions. Capitalizing on the proposed channel model, we analyzed the power gain for both discrete and continuous \ac{IRS}s under narrowband beamfocusing. Our results demonstrate that frequency-dependent beamfocusing is crucial to the successful deployment of future \ac{IRS}-assisted wideband \ac{THz} systems. Regarding future work, it would be interesting to model each IRS element as a transmission line and investigate the wideband design problem. Another promising direction would be to consider the mutual coupling between closely-spaced reflecting elements and its impact on system performance~\cite{irs_coupling}.

\section*{Acknowledgements}
This project has received funding from the European Re- search Council (ERC) under the European Union’s Horizon 2020 research and innovation programme (grant agreement No. 101001331).

\section*{Appendix}
We have that 
\begin{align}\label{integral_approx}
& \frac{\sum_{n=-\frac{N_x}{2}}^{\frac{N_x}{2}-1}\sum_{m=-\frac{N_y}{2}}^{\frac{N_y}{2}-1} e^{-j 2\pi f_s \frac{\tilde{r}_t(n,m) + \tilde{r}_r(n,m)}{c}}L_xL_y}{(N_xL_x)(N_yL_y)}  \nonumber \\
& \approx \frac{\int_{-\frac{\tilde{L}_x}{2}}^{\frac{\tilde{L}_x}{2}} \int_{-\frac{\tilde{L}_y}{2}}^{\frac{\tilde{L}_y}{2}}e^{-j 2\pi f_s\frac{\tilde{r}_t(n,m) + \tilde{r}_r(n,m)}{c}}dxdy}{\tilde{L}_x\tilde{L}_y},\tag{26}
\end{align}
where $\tilde{L}_x = N_xL_x$, $\tilde{L}_y = N_yL_y$, $dx =L_x$, and $dy = L_y$. Now setting $(n-1/2)L_x = x$ and $(m-1/2)L_y = y$ in $\tilde{r}_t(n,m)$ and $\tilde{r}_t(n,m)$, and leveraging the identity
\begin{equation}\label{eq:erf_identity}
\int e^{-jk(ax^2-bx)} dx = \frac{\sqrt{\pi}}{2 \sqrt{jka}}\text{erf}\left(\sqrt{jka}\left(x - \frac{b}{2a}\right)\right)\tag{27}
\end{equation} 
gives the desired result after basic algebra.

\end{document}